\documentclass[a4paper,10pt,twocolumn,abstract]{scrartcl}
 
\usepackage{ifxetex}
\usepackage{ifluatex}
 
\ifxetex
	\usepackage{polyglossia}
	\usepackage{fontspec}
	\usepackage[backend=biber,style=authoryear,language=british]{biblatex}
\else
	\ifluatex
		\usepackage[british]{babel}
		\usepackage{fontspec}
		\usepackage[backend=biber,style=authoryear,language=british]{biblatex}
	\else
		\usepackage[british]{babel}
		\usepackage[style=authoryear,language=british]{biblatex}
	\fi
\fi

\usepackage{csquotes}
\usepackage[continuous]{pagenote}
\usepackage{tikz}
\usepackage{url}
\usepackage[breaklinks]{hyperref}

\ifxetex
	\defaultfontfeatures{Mapping=tex-text}
	\setdefaultlanguage[variant=british]{english}
	\setotherlanguage{dutch}
	\setotherlanguage{german}
\else
	\ifluatex
		\defaultfontfeatures{Mapping=tex-text}
	\fi
	
	\newcommand{\textgerman}[1]{#1}
	\newcommand{\textdutch}[1]{#1}
\fi

\makepagenote

\makeatletter
\renewcommand{\@fnsymbol}[1]{\@Roman{#1}}
\makeatother

\usetikzlibrary{arrows}
\usetikzlibrary{positioning}
\usetikzlibrary{shapes.misc}

\title{Hacktivists:\\Cyberterrorists or Online Activists?}
\subtitle{An Exploration of the Digital Right to Assembly}
\author{J. Slobbe\footnote{\href{mailto:j.slobbe@student.tue.nl}{\nolinkurl{j.slobbe@student.tue.nl}}} \and S.L.C. Verberkt\footnote{\href{mailto:s.l.c.verberkt@student.utwente.nl}{\nolinkurl{s.l.c.verberkt@student.utwente.nl}}}}
\date{4th Junel 2012}

\begin{document}


\maketitle

\begin{abstract}
The last decade, online activism has vastly grown.
In the current digital society, from time to time citizens decide to express their opinion by attacking large corporations digitally in some way.
Where the activists claim this to be a digital assembly, others see it as criminal offences.

In this paper, we will explore the legal and technical borders of the digital right to assembly.
By doing so, we can gain insight into digital manifestations and make up the balance on the digital right to assembly.
As an additional contribution, we will discuss how the digital right to assembly could be granted and which legal and technical requirements should be set for a digital assembly.
\end{abstract}

\section*{Keywords}
Right to Assembly, Digital Assembly, Distributed Denial of Service (DDoS), Virtual Sit-In, Digital Blockade.

\section{Introduction}
Due to groups like ``Anonymous'' and ``LulzSec'', distributed denial of service (DDoS) attacks became a returning topic in newspapers \parencite{Teffer2010a}. 
Not only are the inner workings of a DDoS attack often misunderstood by the general public, the recent avalanche of attacks also showed a new phenomenon: the DDoS attacks were claimed to be an act of activism \parencite{Schouten2010}.

In this paper, we will investigate the use of a DDoS attack as a means of exercising the right of assembly\pagenote{Article 11 of the European Convention on Human Rights}.
Furthermore, we will also discuss how a digital right to assembly could be embodied in current legislation.
For this embodiment, we will base ourselves on the characteristics of the right to assembly and how this relates to the digital world.
We will also present a set of both legal and technical requirements a digital assembly should comply with.

We will start out with discussing the general terminology, in Section~\ref{sec:terminology}, where we will discuss, amongst others, the (D)DoS, (online) activism, and the legal subsidiarity problem.
Afterwards, the legal part of this paper commences with a short survey of the right of assembly, in Section~\ref{sec:rights}.
In this survey, we will discuss notable characteristics of digital availability attacks in the light of the right to assembly.
This is followed by a discussion of attacks on availability in criminal law, in Section~\ref{sec:criminal}, which also notes recent case law on the (D)DoS.
Before the legal section is concluded, in Section~\ref{sec:recap}, we will discuss the policy of the public prosecutor, in Section~\ref{sec:policypp}, and civil disobedience, in Section~\ref{sec:disobedience}.
Finally, we will focus on embodying the right to assembly discussing the option to embed it in the current legislation or to formulate complete new legislation and thereby set requirements to a digital assembly, in Section~\ref{sec:embodying}, followed by a technical adoption, in Section~\ref{sec:technical}.

\subsection{Related work}
According to \textcite{Klang2005}, the current legislation criminalising DDoS attacks in the name of (cyber)terrorism have a serious effect on the civil rights of individuals.
\textcite{Klang2005} vouches for a more modest approach to the DoS technique, which is not build upon the notion of cyberterrorism.

\textcite{Kreimer2001} states that the digital world has the potential of facilitating social movements of all sorts and is increasingly being used for this cause\pagenote{It should be noted that the premature state of digital social movements at the time \textcite{Kreimer2001} was published has been changed to a more mature state.}.
In his paper, besides discussing the value of the digital world for social movements as well as the risks, he also touches on the phenomenon of hacktivism.
The three discussed approaches to hacktivism are digital graffiti, political attacks on systems -- under which he categorises the virtual sit-in, which is discussed in detail in Section~\ref{sec:virtualsitin} --, and the digital release of secret information.
In his argument, \textcite{Kreimer2001} states that the virtual sit-in is probably as legal as repeatedly shouting with the goal of making verbal communication impossible.

\textcite{Samuel2004} constituted a very detailed and broad overview of hacktivism and all its forms.
One of the types distinguished is the performative hacktivism, which she describes as legally ambiguous.
The group of performative hacktivists tries to take a more artistic approach to pressing political issues.
As one of their instruments, the virtual sit-in is named.
\textcite{Samuel2004} notes that the virtual sit-in may be a legal version of the DDoS.

\section{Terminology}
\label{sec:terminology}
Since this paper offers a cross disciplinary perspective on digital assemblies and related subjects, consisting of both computer science and legal studies, we will start with the establishment of terminology.
This way, we can acknowledge the different meanings of terms based on the different fields of science and to prevent us from misunderstandings.
In this section, we will touch upon the phenomena of (distributed) denial of service, (virtual) sit-ins, (online) activism, (cyber)terrorism and other subjects fundamental to this research.

\subsection{(D)DoS}
\label{sec:ddos}
A denial of service (DoS) attack can be defined as an attack on the availability of a system \parencite{Anderson2008}.
A distributed denial of service (DDoS) attack is the same type of attack, but the attack originates from multiple parties.
There are multiple techniques for executing a denial of service attack on a computer system \parencite{Kang2006}. 
Currently, a way to distinguish attacks is based on the method of attack.
The method of flooding boils down to sending lots of data to a server until the maximum capacity is reached and the machine is unable to process new data requests or only able to do so very slowly.
This type of attack is most often executed in a distributed fashion -- possibly using a network of hacked computers --, because it takes lots of data requests for a server to become unavailable.
Notice that this technique is also used for load balance testing -- tests aimed at finding out the maximum number of users a system can take.

Another method of attack is by exploiting known weaknesses in the software used on the server.
Sending a single malicious request to the target could lead to the server software entering a deadlock, crashing, or otherwise being pushed into denying service to other clients. 
One common technique to disable a server that uses a relational database is SQL injection \parencite{Kindy2011}.
SQL Injection is a type of attack on a Web application where the attacker provides SQL code -- the language used to query databases -- as user input to perform unauthorized actions.
This type of attack may allow a single attacker to shut down the service.

However, there are methods that lie in between. 
For example, the DNS amplification attack, as described by \textcite{Vaughn2006}, abuses a property of the DNS infrastructure -- the subsystem responsible for translating domain names to IP-addresses.
This makes it possible for an attacker to generate $8.5$ times more traffic then with an ordinary (D)DoS attack.
Therefore, the critical group of attackers, in case of a distributed attack, is decrease by a proportion $8.5$.

When regarding the physical world, the usage of a method such as DNS amplification can be compared to a group of protesters blocking a bottleneck of a building, e.g. a single entry to the construction.
The impact of this group increases significantly by the use of a bottleneck, when compared to their impact in an open field.
In this paper, we will consider whether this is on the verge of permissible use of weaknesses in an assembly or not.
 
Of course, the discussed techniques can also be combined.
For example, it is possible to use one of the techniques mentioned to attack other services, causing the operating system to fail and making the intended service unavailable.

For average users, some of the techniques may be hard to understand.
However, there exist tools that allow for an attack by means of a simple click on a button.
Such tools are, for example: TFN, TFN2K, Mstream, Naphta, Stacheldracht-V2.66, Stacheldrachtv4, Trinoo, Shaft, IRCbots, FAPI, Targa, Trinity, LOIC\pagenote{\url{http://sourceforge.net/projects/loic/}} \parencite{Dietrich2000}.
For the sake of simplicity, we exclude physical (D)DoS attacks.
In the rest of this paper, the (D)DoS attack is defined as an attack on the availability of an information system from one or more remote systems capable of achieving its goal -- causing unavailability. 

\subsection{Virtual Sit-Ins}
\label{sec:virtualsitin}
The virtual sit-in is when a large group of people rapidly reloads a specific web page \parencite{Samuel2004}. 
This method is popularised by a group of activists that calls themselves the “Electrohippies Collective” \parencite{Klang2005}. 
As opposed to traditional distributed denial of service attacks, this method requires the consent and active participation of every participant.
Thus, the virtual sit-in requires a critical mass in order to be of any effect.

Furthermore, the method of virtual sit-ins does not use the system different from the normal use, besides that the amount of requests is a lot higher than the normal case. 
In other words, the virtual sit-in does not abuse the system in a technical sense, e.g.\ using an exploit.

\subsection{Online Activists}
\textcite{Samuel2004} defines those that non-violently use illegal or legally ambiguous digital tools in pursuit of political goals as hacktivists. 
According to her, hacktivists are a special kind of online activists, that do not necessarily obey order.
In other words, they show civil disobedience in the digital world. 
We can apply this notion of hacktivism to our view on digital activists.
This also encapsulates online civil disobedience.

In the digital world, activists may behave disobedient for political reasons. 
According to \textcite{Klang2005}, these online activists exercise their fundamental rights to freedom of expression\pagenote{Article 10 of the European Convention on Human Rights} and assembly\pagenote{Article 11 of the European Convention on Human Rights}.

To give a bit more context, we will describe some characteristics from two self proclaimed prominent digital activist groups.
As will be explained, these examples do not qualify as digital assemblies.

\subsubsection{Anonymous}
One distinctive characteristic of Anonymous is the fact that it does not have a uniform political opinion with the seemingly exception of the ideal of absolute freedom on the Internet -- e.g. no filtering, censorship or content based payment should be allowed in their opinion.
The communication is decentralized over a broad supply of communication channels, such as image boards, mailing lists and IRC channels.
Since the communication is distributed and the participants are eclectic in their opinions, there is not and cannot be a leader nor spokesmen.
Therefore, every digital protest and no digital protest at all can originate from Anonymous at the same time.

One should always question a press statement by Anonymous, because chances are high that a statement is false -- note that blaming Anonymous is an effective way to masquerade a digital attack.
The general method of Anonymous consists of specifying a target and using the method of DDoS for taking it down, while asking for other participants on the internet. 

\subsubsection{LulzSec}
In comparison to Anonymous, LulzSec focuses on the confidentiality of a system rather then the availability of a system.
LulzSec claims to actively hack systems to publish personal data and credentials stored by governments and multinational corporations.
Self proclaimed members of LulzSec also claim to have delivered important documents to WikiLeaks.

Besides the ideal of absolute freedom on the Internet, these activists share an ideal of transparency of the government and corporations to the society.
It also holds for LulzSec that they do not have a spokesmen or leader.
Therefore, every statement should be treated with suspicion, since the name of the group could be misused to hide another attacker.

Under current legislation, it is clear that the activities of LulzSec are illegal, as they fall within the scope of offences such as intrusion of information systems\pagenote{\textdutch{Artikel 138ab van het Wetboek van Strafrecht}}.
We briefly discussed some characteristics of LulzSec to provide an accurate overview of hacktivism, but will not consider them nor refer to them in the rest of the paper.

\subsection{The Legal Problem of Jurisdiction}
Two important legal problems arise when concerned with legislation in the digital world.
First, we have the subsidiarity legal problem -- or the problem of jurisdiction.
As the digital world is much larger than country borders prescribe, it is difficult to cope with cases where the national legislation of two countries differ.
At the same time, world wide legislation is not easily made -- not even to mention enforcement.
We will not discuss this problem in depth, as it is an entire issue in itself, but it should be noted.

The second problem is the broad spectrum of already existing conventions, which should be taken into account.
In our case, the Budapest Convention on Cybercrime is of importance, which requires the national governments to criminalise the serious hindering of computer systems\pagenote{Article 5 of the Budapest Convention on Cybercrime}.
Of course, this convention complies with fundamental rights and implements related safeguards\pagenote{Article 15 of the Budapest Convention on Cybercrime}.
In Section~\ref{sec:embodying}, we will return to this issue.

\subsection{Technology Change and the Impact on Law}
It is important to have technology neutral legislation, to the extent this is possible.
This means that the legislator may be aware of the technologies at hand, but does not propose legislation that is specifically aimed at a certain technology.
For example, if an expensive DDoS-attack prevention mechanism is invented, legislation that permits specifically DDoS’es as form of assembly would be rendered mostly useless.
However, it may be clear that with multidisciplinary problems, such as those discussed in this paper, the legislator should provide a technical appendix. 


\section{Fundamental Rights and Digital Demonstrations}
\label{sec:rights}
The right of assembly refers to the fundamental right for everyone to have a peaceful assembly, which is granted in the European Convention on Human Rights\pagenote{Article 11 of the European Convention on Human Rights}. 
This charter also grants the related right of freedom of expression\pagenote{Article 10 of the European Convention on Human Rights}.
The right of assembly may only be restricted when it is done lawful, necessary in a democratic society and this is in the interest of national security or public safety, for the prevention of disorder or crime, for the protection of health or morals, or for the protection of the rights and freedoms of others\pagenote{Article 11 member 2 of the European Convention on Human Rights}.

In the Netherlands, the right of assembly is granted in the Dutch Constitution\pagenote{\textdutch{Artikel 9 van de Grondwet}}, as is the right of freedom of expression\pagenote{\textdutch{Artikel 7 van de Grondwet}}. 
Nevertheless, in practice, the rights granted by the European Convention on Human Rights\pagenote{Article 11 of the European Convention on Human Rights and Article 10 of the European Convention on Human Rights} is more important.
This is due to the fact that the courts may not judge acts of parliament and treaties for compliance with the Dutch constitution\pagenote{\textdutch{Artikel 120 van de Grondwet}}.
The right of assembly is regulated in Dutch legislation via the Public Manifestations Act\pagenote{\textdutch{Wet Openbare Manifestaties}} \parencite{Ferdinandusse2001}. 
Restrictions may only be imposed for protection of health or traffic, or to fight or prevent disorder\pagenote{\textdutch{Artikel 2 van de Wet Openbare Manifestaties}}.

It is important to note that the article of the Dutch constitution that grants the right of assembly in its current form, only came into force on 17 February 1988 \parencite{Schilder1989}.
Since that change, the constitution, literally translated, grants the right of assembly and manifestation\pagenote{\textdutch{Artikel 9 lid 1 van de Grondwet}}, where the change boils down to the addition of manifestation\pagenote{\textdutch{Kamerstukken II 1975/76, 13 872, nr. 3 (MvT), p. 38-39}}.
The legislator notes that manifestation refers to a group of people publicly expresses their emotions or wishes concerning a social or political subject\pagenote{\textdutch{Kamerstukken II 1975/76, 13 872, nr. 3 (MvT), p. 39}}.
As, \textcite{Schilder1989} notes, this may be seen as a collective form of the freedom of expression.

Finally, in Germany, the constitution grants the right of peaceful assembly without registration and permission\pagenote{\textgerman{Artikel 8 Absatz 1 des Grundgesetzes}}.
However, this right may be limited when the assembly is “in the open” and the limitation is lawful\pagenote{\textgerman{Artikel 8 Absatz 2 des Grundgesetzes}}.

\subsection{The Blockade as an Assembly}
\textcite{Schilder1989} notes that the Dutch legislator, in response to the question if a blockade may fit within the definition of an assembly, remarked that an action, where the aspect of collective expression has faded in favour of a coercive character towards the government or a third party, cannot be seen as an assembly as meant in the constitution\pagenote{\textdutch{Kamerstukken II 1976/77, 13 872, nr. 7 (MvA), p. 33}}.
However, later the Minister of Internal Affairs added that this does not mean that a manifestation cannot have characteristics of a blockade.
\textcite{Schilder1989} summarises that as long as an assembly in the form of a blockade does not have the goal of imposing a decision or action, it fits the definition.

In 1996, the court of appeal judged in a case between a chlorine producer Solvay and environment defence organisation Greenpeace\pagenote{\textdutch{Hof Amsterdam, 24-4-1996, NJ 1998, 179}}.
Greenpeace had decided to impose a blockade on the chlorine transport of Solvay.
Due to the fact that Greenpeace did not notify Solvay, did not consult the company on the consequences, did not leave room for alternatives, and did not clarify on the duration of their blockade, the intervention with the affairs of Solvay subordinated the interests of the company overly to the interests pursued by Greenpeace.
This resulted into the court ruling that the action of Greenpeace should be qualified as a tort.

Furthermore, the Greenpeace vs Solvay case shows that the freedom of expression\pagenote{Article 10 of the European Convention on Human Rights} may provide the right to infringe other rights\pagenote{\textdutch{Hof Amsterdam, 24-4-1996, NJ 1998, 179}}.
However, due to the complete inconsideration of the interests of Solvay, this argument does not hold.
Similarly, a ruling of the court of appeal in The Hague showed that a blockade may be allowed in the context of a strike under very strict circumstances\pagenote{\textdutch{Hof ‘s-Gravenhage, 22-5-1987, NJ 1988, 646}}.

In another case against Greenpeace, originating from the summer of 1982, the court states that as it is unavoidable that some damages occur during an assembly, this should be allowed to a certain degree\pagenote{\textdutch{Rb. Amsterdam, 26-8-1982, KG 1982, 154}} \parencite{Schilder1989}.
Furthermore, per action, there should be tailored criteria based on to which extent the performing party pursues a general interest, to which extent this could be pursued differently, and the proportionality between the possible damages and the pursued goal of the action\pagenote{\textdutch{Rb. Amsterdam, 26-8-1982, KG 1982, 154 en Rb. Amsterdam, 23-1-1986, KG 1986, 90}}.

Although the additional requirement of peacefulness to the German right of assembly\pagenote{\textgerman{Artikel 8 Absatz 1 des Grundgesetzes}} seems unclear, especially when concerned with blockades, the legislator provides clarification in the German Act on Assemblies and Processions \parencite{Schilder1989}.
This act states that an assembly may only be prohibited in individual cases, where the organising party or his supporters -- not others that want to create havoc at the expense of the assembly -- have a violent or inflammatory goal\pagenote{\textgerman{§ 5 Absatz 3 des Versammlungsgesetz in der Fassung der Bekanntmachung vom 15. November 1978 (BGBl. I S. 1789), das zuletzt durch Artikel 2 des Gesetzes vom 8. Dezember 2008 (BGBl. I S. 2366) geändert worden ist.}}.

The German constitutional court has ruled that blockades -- in this specific case sit-ins -- are not outside the scope of the right of assembly, just because participants are accused of violent coercion\pagenote{\textgerman{§ 240 des Strafgesetzbuch in der Fassung der Bekanntmachung vom 13. November 1998 (BGBl. I S. 3322), das zuletzt durch Artikel 1 des Gesetzes vom 6. Dezember 2011 (BGBl. I S. 2557) geändert worden ist.}}, as the constitutional notion of peacefulness is not as narrow as the broad definition of violence in the law seems to dictate\pagenote{\textgerman{BVerfGE 73, 206 (Sitzblockaden I) und BVerfGE 92, 1 (Sitzblockaden II)}} \parencite{Schilder1989}.
However, six years later, the constitutional court ruled that the term assembly should be interpreted as an orderly gathering of people participating in the public building of opinion, with a focus on discussion or manifestation\pagenote{\textgerman{BVerfGE 104, 92 (Sitzblockaden III)}}.
Furthermore, this ruling decided that it is not up to the parties acting upon their right to assembly to decide how the colliding rights of others may be limited.
Thus, blockades are allowed to that extent where they are the normal side effect of an assembly, and not when its the goal of the assembly to hinder.

\subsection{Assemblies in Privately Held Locations}
If and to what extent rights also have a third-party applicability -- or horizontal function,  which means that they can also hold between two private parties and not only between a private party and the government -- is not stated in the European Convention on Human Rights \parencite{Haeck2005}.
However, it is clear that the European Court of Human Rights does not accept direct third-party applicability \pagenote{Article 34 of the European Convention on Human Rights}.
Nevertheless, as \textcite{Haeck2005} remark, the European Court on Human Rights has accepted cases concerning indirect third-party applicability including cases concerned with the right of assembly\pagenote{ECHR 21-6-1988, no. 10126/82 (Case of Plattform “Ärzte für das Leben” v. Austria)}.
In such a case, a member state gets sued for not protecting the fundamental rights of a citizen against a third party.

The Dutch legislator remarks that fundamental rights may also have a horizontal function\pagenote{\textdutch{Kamerstukken II 1975/76, 13 872, nr. 3 (MvT), p. 15-16}}.
As \textcite{Haeck2005} note, the legislator states that the third-party applicability is to be decided by the judge and not the legislator.
Thus, this is comparable to the European case, where the judge has to balance the interests of the community and those of the relevant parties.

\textcite{Schilder1989} discusses the “Hoog Catherijne”-issue, named after the privately held shopping mall at the central station in Utrecht.
Due to the very public character of the mall, which is highlighted even more by its function as passageway, congress facility, and recreational area, the right of assembly is applicable -- assuming that the fundamental right of assembly is not solely meant for protection against the government, but also has a horizontal function.

It is important that the assembly has a public character, as the Dutch legislator notes\pagenote{\textdutch{Kamerstukken II 1975/76, 13 872, nr. 3 (MvT), p. 39}}.
Of course, it clearly does not make sense to have a manifestation in a location where nobody is able to see it.
The Dutch Public Manifestations Act defines public locations as places that by function or normal usage are open to the public\pagenote{\textdutch{Artikel 1 lid 1 van de Wet openbare manifestaties}}.
As \textcite{Schilder1989} mentions, strikes are a bit different in this respect, as they normally happen at the workplace.
Furthermore, the freedom of choice of location and time of the assembly is deemed important\pagenote{\textdutch{HR, 17-10-2006, AB 2007, 23}} -- this can also be concluded from the ruling of the German constitutional court\pagenote{\textgerman{BVerfGE 104, 92 (Sitzblockaden III)}}.

In conclusion, the public character of an assembly is important for its perceived goal as collective usage of the freedom of expression.
Therefore, a protest in a non-public location does not fit the definition of an assembly in most cases.

\subsection{Relation of DDoS to Fundamental Rights}
Since the European Convention on Human Rights is from 1950, the right to a digital assembly is obviously never granted explicitly.
However, as the medium or infrastructure which is used to execute the right to assembly\pagenote{Article 11 of the European Convention on Human Rights} is never explicitly mentioned, one could argue that an assembly executed over a digital medium, e.g. the Internet, may well be allowed, taking into account the restrictions on the right\pagenote{Article 11 member 2 of the European Convention on Human Rights}.

As a DDoS appears most as a type of blockade, the notes on blockades as a form of assembly are very relevant.
Furthermore, this leads to interpreting virtual sit-in -- not necessarily following the definition of virtual sit-in as proposed by the Electrohippies Collective \parencite{Klang2005} -- also as a type of blockade.
Thus, we will interpret the DDoS -- and the related virtual sit-in -- using the doctrine of blockades and sit-ins as methods of assembly.

By approaching the DDoS as a blockade, we first need to see if it fits in the broad definition of an assembly.
Thus, the character of collective usage of the freedom of expression must be prevalent.
If this is true, we learn that a DDoS may well be protected, albeit with very strict restrictions.
To start, the DDoS may in no way be a method to force a party to a decision or action.
Finally, the interests of the party that is subject to the DDoS must be well-balanced against the goal pursued by the protesting party, thereby taking in account possible different methods to pursue this goal and the proportionality between the damages and the pursued goal.

As discussed before, the public character of an assembly is very important.
Therefore, if a DDoS does only result in downtime, without the public having any idea of the goals the protesters pursue, it cannot be deemed an assembly.

\subsection{Individual versus Collective Actions}

As discussed before, the Dutch constitution recognises the right to assembly and manifestation\pagenote{\textdutch{Artikel 9 van de Grondwet}}.
However, it does not make a distinction between individual and collective assemblies -- being computer scientists, we consider an empty set still as a set.
Nevertheless, the Dutch legislator gives collective meaning to the definition of assembly\pagenote{\textdutch{Kamerstukken II 1975/76, 13 872, nr. 3 (MvT), p. 39, Kamerstukken II 1985/86, 19 427, nr. 3 (MvT) en Kamerstukken II 1985/86, 19 427, nr. 5 (MvA), p. 8.}}. 
Furthermore, the European Convention on Human Rights does not make this distinction either\pagenote{Article 11 of the European Convention on Human Rights}, nor does it distinguish between the right of assembly and collective freedom of expression\pagenote{Article 12 of the European Convention on Human Rights}.

The Dutch Public Manifestations Act does not provide direct clarification on the difference between individual and collective actions\pagenote{\textdutch{Artikel 1 van de Wet Openbare Manifestaties}}.
However, this act does regulate that every municipality could have a different implementation strategy of the act in local regulations\pagenote{\textdutch{Artikel 4 van de Wet Openbare Manifestaties}}.
Therefore, it is possible that one municipality makes a distinction between collective and individual protest and the other does not.

In practice, according to \textcite{Embregts2007}, there is a difference in the right to protest as individual versus the right to protest as a collective.
As an individual protests, this is not a case for the right of assembly, but for the right of free speech\pagenote{\textdutch{Artikel 7 van de Grondwet}}.
Therefore, the Public Manifestations Act is not applicable.
There are two exceptions.
First, when other individuals spontaneously join the protest it becomes an assembly.
The second exception is when the individual protest is during another protest.

In conclusion, a strictly individual manifestation is protected by the freedom of speech and is not considered an assembly.
Therefore, individual actions are even better protected by fundamental rights than collective actions, as there are more safeguards implemented.


\subsection{Boundaries on the Fundamental Rights of assembly}
\label{sec:boundaries}
In 2007, the Dutch ombudsman released a report in which the boundaries for a physical demonstration as described by the Dutch public manifestations act\pagenote{\textdutch{Wet Openbare Manifestaties}} are clarified \parencite{Embregts2007}.
The ombudsman states that, although protesting is a fundamental right, this right may be infringed if the protest disproportionally disrupts security, health or property.
The boundaries given by the ombudsman could be a good base for testing the digital protest.
The first important boundary is that there is a registration period of 4 days for a collective protest.
This makes it possible for the object protested against to enforce an injunction if it thinks the protest disproportionally harms his interest.

The second interesting point is that the identification duty\pagenote{\textdutch{Artikel 447e van het Wetboek van Strafrecht}} is widened \parencite{Embregts2007}.
The protester does not have to be suspected of a criminal offence, but could be enforced to identify himself as part of the protection of order -- even if the police already knows the suspect.
For regulating purposes, the law enforcement authorities appreciate a scheme from the organisation about the way the protest is organised.
This, however, is not an obligation.


A third interesting aspect is that intervention of the police when the protest is near an (important) government building is allowed \parencite{Embregts2007}.
A restriction is that it is forbidden to use tools which could be used as weapons -- such as sticks -- during the protest.
This restriction could have interesting implication in a digital protest.
The last obligation to the protest is that the protesters have to leave the location of protest clean after the action.
This could also have interesting implication for a digital protest.
Finally, the government has a department for facilitating manifestations and assemblies.

%
%

\section{Criminal Law and Availability Attacks}
\label{sec:criminal}
The Budapest Convention on Cybercrime, which entered into force in 2004 and was enacted in the Netherlands in 2007, urges the national governments to criminalise the intentionally and without right hindering the functioning of a computer system inputting, transmitting, damaging, deleting, deteriorating, altering or suppressing computer data\pagenote{Article 5 of the Budapest Convention on Cybercrime}.
At the time of ratification, most Dutch law was already in line with the convention\pagenote{\textdutch{Kamerstukken II 2004/05, 30 036, nr. 3, p. 1}}, part of the pending act “Computercriminaliteit II”\pagenote{\textdutch{Kamerstukken II 1998/99, 26 671, nr. 1-2}} -- either originally or as an extension due to the convention.
As the name suggests, at this time, there already was an act called “Computercriminaliteit”\pagenote{\textdutch{Kamerstukken II 1989/90, 21 551, nr. 1-2}}, which, amongst others, contained the first criminalisations of attacks on the availability of automated systems.

Dutch law has several criminalisations of attacks on the availability of automated systems.
To start, the intentionally rendering unavailable of an system for telecommunications is punishable with up to a one year sentence if this affects the transmission, storage or processing of data within a public telecommunication network or a public telecommunication service; punishable with a sentence of up to six years if this damages goods or service; punishable with a sentence of at most nine years if this endangers life; and punishable with a sentence of up to fifteen years if this led to death\pagenote{\textdutch{Artikel 161sexies lid 1 van het Wetboek van Strafrecht}}.
If the act was unintentional, sentences drop to at most six months in the first two cases, up to one year in the third case, and at most two years imprisonment in the last case\pagenote{\textdutch{Artikel 161septies van het Wetboek van Strafrecht}}.
The legislator notes that these articles also criminalise the creation of unavailability by sending a lot of data to some electronic address\pagenote{\textdutch{Kamerstukken II 1996/97, 25 533, nr. 3 (MvT), p. 69 en Hof 's-Gravenhage, 10-2-2006, LJN AV1452}}.

The two offences discussed in the previous paragraph where introduced as the modern society became more and more dependent on automated systems\pagenote{\textdutch{Kamerstukken II 1989/90, 21 551, nr. 3 (MvT)}}.
The legislator even thought computer networks to be so important that a strike was not accepted as an exception to the offences.
Nevertheless, it is later clarified by the Dutch legislator that this article is only concerned with automated systems of general interest\pagenote{Kamerstukken II 1990/91, 21 551, nr. 6 (MvA), p. 13}.
Furthermore, it is also noted that it is not focussed on cases where there only is slight interference and no real damage to the services.

Furthermore, it is not allowed to intentionally and without right make data unavailable that is stored using an automated system\pagenote{\textdutch{Artikel 350a van het Wetboek van Strafrecht}}.
If this is done with hacking the sentence goes up to four years\pagenote{\textdutch{Artikel 350a lid 2 van het Wetboek van Strafrecht}}, and without up to two years\pagenote{\textdutch{Artikel 350a lid 1 van het Wetboek van Strafrecht}}.

Finally, a sentence of at most one year can be given for the intentionally and without right making unavailable of an automated system by sending or offering it data\pagenote{\textdutch{Artikel 138b van het Wetboek van Strafrecht}}.
One of the main motivators for this article was spamming\pagenote{\textdutch{Kamerstukken II 1998/99, 26 671, nr. 3 (MvT)}}.


\subsection{Case law}
In 2005, the local court in The Hague decided that a DDoS attack may well be punished under the conditions of the first part of the first member of article 161sexies of the Dutch penal code\pagenote{\textdutch{Artikel 161sexies lid 1 sub 1 van het Wetboek van Strafrecht}}, as it overloads the network thereby affecting the complete automated work\pagenote{\textdutch{Rb. 's-Gravenhage, 14-3-2005, Computerrecht 2005, 36, Rb. 's-Gravenhage, 14-3-2005, LJN AT0222, Rb. 's-Gravenhage, 14-3-2005, LJN AT0224, Rb. 's-Gravenhage, 14-3-2005, LJN AT0230, Rb. 's-Gravenhage, 14-3-2005, LJN AT0239 en Rb. 's-Gravenhage, 14-3-2005, LJN AT0249}}.
Furthermore, in this specific case one website was attacked, but the attack resulted in downtime at other clients of the internet service provider that hosted the attacked website, which caused the judge to finding the second part of the article also applicable, as the services offered where broadly interrupted\pagenote{\textdutch{Artikel 161sexies lid 1 sub 2 van het Wetboek van Strafrecht}}.

In the higher appeal of these cases was referred back to the comment of the legislator that the articles the conviction was based on where relevant\pagenote{\textdutch{Kamerstukken II 1996/97, 25 533, nr. 3 (MvT), p. 69}}, even though the article of the Dutch penal code specifically meant for DDoS attacks\pagenote{\textdutch{Artikel 138b van het Wetboek van Strafrecht}} was still pending in parliament\pagenote{\textdutch{Hof 's-Gravenhage, 10-2-2006, LJN AV1452}}.
The other higher appeals had a comparable outcome\pagenote{\textdutch{Hof 's-Gravenhage, 10-2-2006, LJN AV1449, Hof 's-Gravenhage, 10-2-2006, LJN AV1451 en Hof 's-Gravenhage, 10-2-2006, LJN AV1454}}.

However, another case from 2005\pagenote{\textdutch{Rb. Breda, 10-11-2005, LJN AU6703}} led to acquittal for the court of appeal\pagenote{\textdutch{Hof 's-Hertogenbosch, 12-2-2007, LJN BA1891}}.
The main motivation was that the court decided it to be not proven that the network of the given internet service provider was endangered, only the specific system, which made the court decide that article 161sexies of the Dutch penal code\pagenote{\textdutch{Artikel 161sexies van het Wetboek van Strafrecht}} was not applicable.

From the case law, we can learn that in order to be convicted on the basis of article 161sexies of the Dutch penal code\pagenote{\textdutch{Artikel 161sexies van het Wetboek van Strafrecht}}, at least an automated system of network general interest needs to be endangered\pagenote{\textdutch{HR, 22-2-2011, RvdW 2011, 317, Hof 's-Hertogenbosch, 12-2-2007, NJFS 2007, 184 en Hof 's-Hertogenbosch, 12-9-2008, NJFS 2008, 212, LJN BF0770}}.
Nevertheless, in cases where other websites beside the website that is subject of the ddos experience downtime, the article is applicable again\pagenote{\textdutch{Rb. Rotterdam, 14-4-2010, NJFS 2010, 173, LJN BM1172}}.




\subsection{Grounds of Justification}
\textcite{Mevis2009} notes that the Dutch Penal Code provides two types of grounds for exclusion of criminal liability. 
Namely, grounds for justification -- grounds which regard the wrongfulness of the act -- and grounds for excuse -- grounds which regard the culpability of the actor.
Of these, only the first type is of interest to our research, as we are not interested in personal circumstances, but only in the general act.

Especially Force Majeur\pagenote{\textdutch{Artikel 40 van het Wetboek van Strafrecht}} is of relevance.
This article provides, due to the case law, both a ground for justification as a ground for excuse \parencite{Mevis2009}.
The ground of justification created by this article, also known as “state of emergency”, justifies punishable behaviour when there is a conflict between the duty to solve an emergency and the duty to comply with the law -- given that the emergency is substantial in proportion to the criminalisation\pagenote{\textdutch{HR, 16-9-2008, LJN BC7938}}.
\textcite{Mevis2009} illustrates this type of Force Majeur using a case where an optician broke the law by opening his store outside of the allowed shopping hours in order to provide a client in dire need with a pair of glasses\pagenote{\textdutch{HR, 16-10-1923, NJ 1923, 1329}}.

In some cases, the description of an offence contains the requirement that it happened without right \parencite{Mevis2009}.
The legislator normally adds this requirement when the described offence also frequently happens rightfully.
The addition of “without right” requires the public prosecutor to explicitly prove that the action was without right, whereas if this is omitted, the action is always considered to be without right, unless the defence submits a ground for justification.
In other words, the burden of proof with regard to justificatory grounds is changed by the addition of “without right”.
For example, the description of the offence destruction\pagenote{\textdutch{Artikel 350 van het Wetboek van Strafrecht}} contains the requirement “without right”, which, amongst others, clears honest demolition companies of the possible criminal charge of destruction.

As discussed before, one of the offences commonly charged in cases concerning a form of DDoS does not require the offence to be without right\pagenote{\textdutch{Artikel 161sexies van het Wetboek van Strafrecht en Artikel 161septies van het Wetboek van Strafrecht}}.
Furthermore, the legislator claims to have omitted this requirement due to the dependability on automated systems and networks in our modern society\pagenote{\textdutch{Kamerstukken II 1989/90, 21 551, nr. 3 (MvT), p. 20 en Kamerstukken II 1990/91, 21 551, nr. 6 (MvA), p. 35-36}}.

\textcite{Mevis2009} also mentions absence of substantive wrongfulness, which is an implicit ground for justification.
This provides a ground of justification when the act perfectly fits the description of the offence, but the offender objectively acted in favour of the cause the criminalisation tries to pursue\pagenote{\textdutch{HR, 20-2-1933, NJ 1933, 918}}.
However, this does not fit idealistic causes\pagenote{\textdutch{Rb. Haarlem, 5-6-2008, LJN BD5505}}.
Therefore, it is not an applicable ground of justification in the present research.

\section{Policy of the Public Prosecutor}
\label{sec:policypp}
As \textcite{Mevis2009} discusses, the public prosecutor may choose not to prosecute\pagenote{\textdutch{Artikel 167 van het Wetboek van Strafvordering en Artikel 242 van het Wetboek van Strafvordering}}.
In practice, there is the possibility of technical dismissal -- when it is not likely the prosecution will lead to a conviction -- and the dismissal on the basis of the expediency principle\pagenote{In Dutch: \textdutch{opportuniteitsbeginsel}} -- when the public prosecutor decides prosecution is not opportune.

When the public prosecutor considers a conviction possible, but thinks prosecution is not in favour of the general interest, he may decide to a dismissal out of opportunity, as he considers prosecution not to be opportune \parencite{Mevis2009}.
This is due to the fact that Dutch criminal law is based on the principle of opportunity, as opposed to the principle of legality -- as German criminal law is.
Such a decision that prosecution is not opportune is a decision of policy.
Furthermore, the public prosecutor may create guidelines, which describe when he will not prosecute.
Thus, a decision to tolerate a certain crime may be formulated in the form of such a guideline -- a common example in Dutch criminal law is the policy on soft drugs\pagenote{\textdutch{Aanwijzing Opiumwet (2011A021)}} and euthanasia\pagenote{\textdutch{Aanwijzing vervolgbeslissing inzake levensbeëndiging op verzoek (euthanasie en hulp bij zelfdoding) (2006A009 )}}.

Recently, the Public Prosecutor ordered two suspects of performing a DDoS on the websites of MasterCard and VISA to community service\pagenote{\url{http://www.om.nl/actueel-0/nieuws-persberichten/@157968/taakstraffen/}}.
These actions were -- allegedly -- performed with political reasons, namely the blocking of donations to WikiLeaks by these credit card companies \parencite{Teffer2010a,Teffer2010}.
The public prosecutor decided not to prosecute the suspects under one of the offences discussed in Section~\ref{sec:criminal} -- most notably 161sexies of the Dutch penal code\pagenote{\textdutch{Artikel 161sexies van het Wetboek van Strafrecht}} --, but use their competence to give a punishment order\pagenote{\textdutch{Artikel 257a van het Wetboek van Strafvordering}} -- which is allowed for offences penalised with up to 6 years of detention.
The underage offender was ordered to 26 hours of community service, whereas the older suspect was ordered to 80 hours of community service.

\section{Civil Disobedience}
\label{sec:disobedience}
After discussing the (D)DoS method with respect to the European rights, criminal law and the policy of the public prosecutor we decided
to analyse the (D)DoS method with respect to civil disobedience to harvest (more) requirements which could bind a legal requirement.

With civil disobedience, we mean events in which citizens disobey the law, claiming to do so for a greater good \parencite{Klang2005}. 
An example of civil disobedience can be found in Rosa Parks, who, in 1955, refused to give up her seat in the bus to a white man, whereby she broke the law. 

She performed this action, to oppose against discrimination and thereby breaking the dogma of racial differences.

In this section the guidelines provided by \textcite{Schuyt1972} will be discussed in the context of digital civil disobedience.
His first guideline prescribes that the violation of law should be the result of a process of deliberation.
Therefore, the time and aim of the action may not have a spontaneous character.
This guideline could be adopted in the digital infrastructure without the loss of generality.

The second guideline of \textcite{Schuyt1972} describes that there must be a relation between the method chosen to protest and the objective of the protest.
Therefore, if the objective of protest concerns the physical infrastructure, then a digital protest is not eligible until a later stage, when the physical options are exhausted.
However, if the objective concerns the digital infrastructure, then a digital protest is appropriate.

The third guideline prescribes public violation of the law \parencite{Schuyt1972}.
If the protester does not try to hide his actions, he stands stronger in influencing the public opinion.
A typical DDoS attack could be anonymous and mere downtime does not express much opinion.
Therefore, this guideline needs technical adaptation when imposed on DDoS as an assembly.
Furthermore, this also fits with the public property that the Dutch legislator notes of assemblies\pagenote{\textdutch{Kamerstukken II 1975/76, 13 872, nr. 3 (MvT), p. 39}}.

Besides the requirement that the violation should have a public character, \textcite{Schuyt1972} also requires voluntary cooperation when prosecuted.
Therefore, the disobedient protester should not use a proxy or other means of hiding and has to store (forensic) evidence to be able to cooperate with possible prosecution.

As a fifth requirement, \textcite{Schuyt1972} notes the absence of violence.
Although the technique of DDoS is classified as an attack, it does not necessarily cause physical harm to individuals or property -- although it is possible.
However, if the technique is used against critical systems as, for example, SCADA systems \parencite{Jidong2008}, it could create major physical damage.
Since the protester is not always able to decide whether a system he blocks is important -- for example, when a website is run on the same mainframe as a critical system --, the protester has the responsibility to exclude this risk.
In Section~\ref{sec:embodying}, we will discuss both technical and juridical requirements to mitigate this risk.

A long manifestation could harm the interests of other citizens, which could violate the sixth requirement: the rights of others citizens should be respected as much as possible.
Nevertheless, this, again, boils down to an act of balancing the interests of all involved parties.

The seventh requirement from \textcite{Schuyt1972} prescribes that all legal resources need to be exhausted.
One should not directly resort to civil disobedience, but first seek to use other resources such as influencing the public opinion, trying to enforce your rights in court, and the likes.
This requirement does not differ from the physical world, because most of the legal resources are constituted in the physical world.

The disobedient citizen has to accept the risk of being punished \parencite{Schuyt1972}.
For common violations (e.g. entering private terrain or temporary occupation) the maximal punishment could be acceptable to the disobedient.
However, the punishment for a DDoS could be 6 years.

Finally, the disobedient citizens is acting as he does to provoke a trial case in court, in order to test his moral grounds of justifications for a judge.

\section{Recap of Current Legislation}
\label{sec:recap}
To get a good picture of the status quo of digital manifestation in Dutch and European legislation, we performed a thorough legal study.
At this point, we will summarise the results of the legal part of the research and give a brief description of the current status of the digital right to assembly.

In Section~\ref{sec:rights}, we have seen under which conditions the DDoS may be considered as a method of exercising the right to assembly.
To start, the DDoS has to fit the description of an assembly, which requires it to have a public character and to have a prevalent character of collective expression of opinion \emph{\textbf{(definition)}}.
Furthermore, the DDoS can be seen as a virtual blockade -- and, thus, virtual sit-in.
This gives the following requirements for a DDoS to be considered as covered by the right to assembly: reasonable alternatives have already been pursued \emph{\textbf{(subsidiarity)}}, the (possible) damages of the manifestation are proportional to the interests of the party subject to the protest \emph{\textbf{(proportionality)}}, and the protest is in the general interest \emph{\textbf{(necessary)}}.
As we have also seen, in Germany it would be required that the blocking effect of the DDoS is a side effect of a normal peaceful assembly, which will commonly not be the case.

Especially the requirement of visibility makes it hard for a DDoS to be deemed a reasonable blockade, as this would require artificial interventions, such as the use of social networking, microblogs or comparable services, where the protesting citizens unite.
As the activity should have a character of publicly collectively exercising the freedom of expression and not of hindering services, a loosely connecting campaign on a social networking site could be seen as the assembly, but the DDoS itself would still be questionable.
However, their may be variants possible that withstand this requirement, such as flooding an email-address or contact form with the message of the protest.
Please note that requirements considering the balancing of interest or the pursuing of the goal through other means make it less likely for a DDoS to be considered as means of exercising the right to assembly in practice, they do not make the change for that to happen negligible.

As we saw in Section~\ref{sec:criminal}, the DDoS is criminalised in Dutch legislation.
However, it is not completely unambiguous which offences in particular fit the bill in general, which leaves this task to case law.
Due to the omission, of “without right” in the most commonly used offence, any action that fits the description is considered to be without right.
In combination with the importance the Dutch legislator has in mind, when it comes down to information networks, it seems unlikely the court will allow an appeal to necessity as a ground of justification using the right to assembly in a DDoS case.

Therefore, we can safely assume that, in most cases, it is not possible to call upon the right to assembly for a DDoS attack.
Furthermore, there currently is no policy of the public prosecutor considering digital rights, which we discussed in Section~\ref{sec:policypp}.
Nevertheless, the door is still open to the in Section~\ref{sec:disobedience} discussed phenomenon of civil disobedience.

\section{Embodying the Digital Right to Assembly in Legislation}
\label{sec:embodying}
For the next step in this research, we will ask ourselves to what extent digital assemblies should be possible and how this should be embodied in the current legislation. 
As the first question -- whether the digital assembly should be possible -- is mostly a political question, we will focus on the second question.
In order to do this, we will first evaluate if this requires extending or creating fundamental rights, followed by the possibility of the public prosecutor issuing a special policy.

\subsection{Extending Current Rights}
In the Dutch constitution, the secrecy of correspondence is granted\pagenote{\textdutch{Artikel 13 van de Grondwet}} \parencite{Asscher1999}.
As this only protects written letters, telegraphs, and telephone conversations, it is not of use for other forms of communication.
However, nowadays, the right to privacy\pagenote{\textdutch{Artikel 10 van de Grondwet}} is understood as also including private communication.
Thus, privacy was extended to fill the gap of the secrecy of correspondence of other forms of communication.
In analogy to this example, we will evaluate whether an extension of the existing fundamental rights is feasible.

As discussed in Section~\ref{sec:rights}, the Dutch constitution grants the right to assembly and manifestation\pagenote{\textdutch{Artikel 9 lid 1 van de Grondwet}}, as opposed to the European Convention on Human Rights, which only grants the right to assembly\pagenote{Article 11 of the European Convention on Human Rights}.
Merely the addition of a “right to manifestation” to the right to assembly may enable digital assemblies under strict conditions -- that fit in with the requirements we will discuss in Section~\ref{sec:requirements}.
Furthermore, this solution clarifies the general place of the manifestation, which lies somewhere between the right to assembly\pagenote{Article 11 of the European Convention on Human Rights}, the right to freedom of expression\pagenote{Article 10 of the European Convention on Human Rights}, and the right to association\pagenote{Article 11 of the European Convention on Human Rights}.

There are two noteworthy risks with extending existing rights.
Firstly, most rights form a foundation for other legislation that builds on it, which leads to difficulties when those rights are changed.
Secondly, it can feel as if the legislator tried very hard to fit in additional rights with current legislation, when there actually are enough differences to justify introducing a new right.

\subsection{Creating New Rights}
When the Dutch government found that intrusion into information systems was not punishable under trespassing, the legislator decided to create a new offence for intrusion into information systems\pagenote{\textdutch{Kamerstukken II 1989/90, 21 551, nr. 3 (MvT)}}.
The new offence is based upon the offence of trespassing, but is also an adaptation to the digital world.
With this example in mind, we will discuss the creation of a new fundamental right to enable digital assemblies.

The discussed “right to manifestation” could also be introduced as a separate right, which has some benefits.
Most notably, this makes the right truly separate of previous rights, and does not give it the position of little brother in comparison to the right to assembly.
Furthermore, this leaves room for a more specific approach to the right and its safe guards.

The most notable backside to introducing new rights is that this eventually leads to devaluation due to the existence of too much fundamental rights.
However, it seems that the right of manifestation is recognised highly, and thus not capable of causing such devaluation.

\subsection{Policy of the Public Prosecutor}
As we discussed in Section~\ref{sec:policypp}, it may be possible to protect the digital assembly by policy of the public prosecutor.
As policies of the public prosecutor are a typical Dutch phenomenon, this approach is not very useful in the global issues generated by the digital world.
However, it may be of use as a temporary solution.


\section{Requirements for Digital Assemblies}
\label{sec:requirements}
A digital assembly should comply with a number of requirements.
To find these requirements, we will look back at the general doctrine on assemblies and try to adopt this to the general world.
In addition, we will consider differences between the digital and the physical world and try to find requirements that settle these differences.

In Appendix~\ref{sec:relations}, the relations of the several proposed requirements to each other and the general doctrine are displayed graphically.
These figures are referred for a more detailed overview of the requirements.

\subsection{Defining Digital Assemblies}
We have seen that the Dutch legislator considers a manifestation as a group of people that publicly expresses their emotions or wishes concerning a social or political subject\pagenote{\textdutch{Kamerstukken II 1975/76, 13 872, nr. 3 (MvT), p. 39}} and that the German constitutional court decided that an assembly should be interpreted as an orderly gathering of people participating in the public building of opinion, with a focus on discussion or manifestation\pagenote{\textgerman{BVerfGE 73, 206 (Sitzblockaden I), BVerfGE 92, 1 (Sitzblockaden II) und BVerfGE 104, 92 (Sitzblockaden III)}}
Our idea of a digital assembly will build further upon these perceptions.
Thus, we will consider visibility, the expression of opinion, and collectivity as requirements.

\subsubsection{Visibility}
\label{sec:reqvisible}
That assemblies should be visible is one of the major requirements to an assembly, as this can make the difference between an action where the goal of hindering is prevalent and an action where the goal of publicly expressing an opinion prevails.
The same goes for a digital assembly.
Thus, the public should be able to see the opinion that the assembly tries to express.

\subsubsection{Expression of Opinion}
\label{sec:reqopinion}
Besides being public, in an assembly the goal of collectively expressing an opinion should prevail.
This means that a digital assembly should be politically or socially motivated, and not be aimed at, for example, smear campaigns\pagenote{\textdutch{Artikel 262 van het Wetboek van Strafrecht}}.

\subsubsection{Collectivity}
\label{sec:reqcollective}
The collective character of an assembly is one of the reasons of its power.
If a very large group of citizens opposes a certain decision, they are more likely to be heard, than if an individual does.

However, in the digital world, automation is cheap.
Thus, one could act as if he is a group of people.
Therefore, the requirement of collectivity prescribes the “one man, one vote” principle.

Furthermore, a certain proportionality between the size of the target and the size of the participants in the assembly and the impact of the assembly is important.
In other words, assemblies require a certain critical mass, in order to be of effect.

\subsection{Permissibility of Digital Assemblies}
The court evaluates the permissibility of an assembly on several conditions.
First of all, the action may not be a method to force some party to a decision or action -- please note that this is also in line with the general requirement that the expression of opinion should be a prevalent characteristic.
Secondly, the goal pursued by the protesting party should be well-balanced against the interests of the other party, especially concerning possible damages.
Finally, other possible methods to pursue the goal should have been taken into account or tried.

\subsubsection{No Coercion}
\label{sec:reqnocoercion}
As stated before, the non-permissibility of coercion lies also with the requirement that an assembly is mainly about the expression of opinion and not about hindering or comparable goals.
Of course, there is a slight difference, as the assembling party may try to force the opposing party with social pressure to follow the opinion they are expressing. 
However, forcing them at gunpoint is obviously not permissible at all, and can be considered a form of coercion.

\subsubsection{Proportionality}
\label{sec:reqproportionality}
As we have seen before, proportionality is always an important consideration.
In this case, it is of great importance that there is a balance between the goal that is pursued and the interests of the party that is subject to the protest.
For example, if a protest aims at a local milk factory which actually is bound by the decisions of the municipality, the protest may not be proportional.
However, if it was their own decision that caused the protest, it may be proportional.

The same reasoning could be used for damages.
A small wrong does not justify a lot of damages, but a large issue may.
Proportionality between the pursued goal and the potential damage to the interests of the opposing party is, thus, an important requirement.
For this, the practicalities can be left to the judge.

\subsubsection{Subsidiarity}
\label{sec:reqsubsidiarity}
An assembly -- especially a more intrusive one -- should probably not be the first answer when someone feels their interests -- or opinions -- are not heard.
It is important that those that want to express their opinion have followed other possible paths in the pursue of their goal, before they step up to more heavier methods, such as the assembly.
Therefore, it is important to evaluate whether other methods of reaching the intended goals are sufficiently exhausted.


\subsection{Enabling Orderly Digital Assemblies}
\label{sec:orderly}
Finally, there are requirements which make digital assemblies orderly.
These requirements are of a lower importance, as they are not as much build on the legality of the digital assembly, but say something about the ability to keep order.
This is not only beneficial for the authorities, but also for the protesters, as they have less chance of their assembly getting hijacked for violent or otherwise unwanted reasons.

In practice, the municipality regulates the held manifestations \parencite{Embregts2007}.
Although most of the regulations by the municipality are not strictly required, protesters are urged to comply with them, to prevent excesses.
This leads to a central organisation for each assembly, which also announces the activity.
Furthermore, the police will provide supervision, which will take action when individuals step beyond the law.


\subsubsection{Supervision}
\label{sec:reqsupervision}
Within a physical demonstration, the presence of law enforcement officers is a preventive measure against escalation.
The officers can inform the mayor about the status of the protest, providing him with information to make a decision whether the protest should be broken up.

This requirement is hard to satisfy, since the anonymous nature of the Internet prevents it, although it could be satisfied when part of the infrastructure is provided by the government.
Nevertheless, chances are fairly high that this infrastructure raises distrust and may for that reason not be used.

\subsubsection{Central Organisation}
\label{sec:reqcentral}
Most of the larger manifestations have a central organisation.
This organisation registers the manifestation with the municipality and tries to help the law enforcement to keep the organisation orderly.
However, in Section~\ref{sec:boundaries}, we saw that it is not required to register a manifestation nor to have a central organisation.

Even a digital assembly requires some sort of announcement to gather those who want to protest at the right time, for the protest to be successful.
After this action, the protest can be considered registered.
In addition, details of the manifestation could be provided using some sort of pseudonymity or anonymity.

\subsubsection{Announcement}
\label{sec:reqannouncement}
Commonly, the municipality asks its citizens to register manifestations.
Although this is not always required -- for example, ad hoc manifestations cannot be registered --, this is an interesting requirement to discuss, due to the risk of misconfiguration of a system, which could lead to much extremer effects than expected in a normal situation.

An announcement should be unambiguous and provably delivered to the legal subject who is subject to the manifestation.
This gives the legal subject of protest the option to ask for a delay or rejection by filing an injunction in court.
The time granted by the judge should be sufficient to prevent great disaster. 
However, it should not be sufficient to disarm the protesting party.
For example, if the subject of the protest does not repair the misconfiguration, the protest will be allowed at some point.
Please note that, in this example, this badly configured server is targeted, not the misconfiguration itself.

\section{Technical Requirements for Digital Assemblies}
\label{sec:technical}
In this section, we will discuss the technical requirements, based on the legal requirements, to make digital assemblies possible.
Based on the requirement of visibility, we will introduced a technical requirement concerning visibility.
The requirement of collectivity brings us to introduce one man one vote, and group proportionality as technical requirements.

We will also consider additional requirements that provide for an orderly assembly.
From the legal requirement of supervision, we come to the use of revocable anonymity.
Finally, although we noticed the existence of central organisations in Section~\ref{sec:reqcentral}, technically we acknowledge the importance of a decentralised solution.


\subsection{Visibility}
\label{sec:techvisible}

A concrete example, which satisfies the visibility requirement, is protesting by e-mail. 
If a digital assembly is shaped in the form of sending multiple e-mails at a predefined time as a group, it will have the properties of an attack on the availability of a system.
Furthermore, if enough individuals participate in sending e-mails with, for example, large attachments and prevention against technical countermeasures, such as spam filters, the content is clearly visible to the subject of the protest.

The visibility becomes even stronger if carbon copies are used, in order to send a copy of the e-mail to a public place, where it is for everyone to see \parencite{rfc5322}.
The i-box of the party that is subject to the protest may become so full that it is not feasible to use at that moment.
Even stronger, the memory of the mail server could fill up and cause the server to crash, which is a (D)DoS by definition \parencite{Bass1998}.
A recent example of such a protest over e-mail is when a Dutch member of the European Parliament placed a call on citizens to send protest e-mails against ACTA to all the members  of the European Parliament\pagenote{\url{http://www.reddit.com/r/politics/comments/ow1v5/acta_note_from_marietje_schaake_member_of_the/}}.

Within a physical demonstration, the ground where the protesters assemble is impervious for other people. 
However, when they watch the assembly from above -- e.g. from a building -- or from the side, they can clearly see the statements of the group.
Beside carbon copying protest e-mails to publicly accessible boards, social media could get filled with the collectively expressed opinion of the manifestation.

\subsection{One Man, One Vote} 
\label{sec:techonemanonevote}
The infrastructure must ensure that each person has at most one vote -- e.g. one device -- in the manifestation. 
In the example of one device, this device may not be able to generate more than the average of possible request, i.e. super computers are not allowed.

This requirement mainly ensures that the use of bot-networks \parencite{Rajab2006} or hired server park capacity does not become legal for the use of protesting. 
It also contributes to the requirement of group proportionality on which we will elaborate in Section~\ref{sec:techgroupproportionality}.

\subsection{Group Proportionality} 
\label{sec:techgroupproportionality}
The impact of the assembly should be proportional to the size of the target and the size of the group of participants to the assembly.
Therefore, it should not be possible for one person to take down another party single handed -- e.g. by exploiting a vulnerability.
Comparably, a small group of protesters should not be able to take down a corporation that is much larger.
Thus, the used techniques should comply with this requirement of group proportionality.

 
\subsection{Revocable Anonymity}
\label{sec:techrevocable}
An important requirement for a protest is that the protesters are anonymous.
However, when one of the protesters abuse the anonymity by, for example, trying to hack the server, the investigative authorities should be able to track this individual down. 
Since these authorities should not be able to track the whole group down or otherwise abuse this procedure, revocable anonymity can be put in place \parencite{Koepsell2006}. 

For revocable anonymity to work, some entity needs to be in charge of revocation.
This can be implemented in multiple ways.
For example, critical mass based revocation could be used, where the protesters give up on people trying to disturb the event.

\subsection{Decentralised}
\label{sec:techdecentral}
As a non-negligible amount of the manifestations is aimed at the government, people may not have trust in a infrastructure which is actively maintained or supplied by that same government.
Therefore, it may not be effective for a government to straightforwardly implement these requirements.

Furthermore, as decentralisation is in the very veins of the Internet, the digital assembly infrastructure should be as decentralised as possible, thereby omitting central contributors to the infrastructure, such as the government.
In addition, the government should only contribute to parts that are necessary for law enforcement in the case a manifestation escalates.

\section{Conclusion}
%

In this paper, we have investigated the digital right of assembly, especially considering attacks on the availability of information systems as assembly, e.g. the DDoS.
This lead us to conclude that, in most cases, such attacks cannot be deemed a means of exercising the right to assembly.
This is due to the fact that such attacks, which can be seen as digital blockades, do commonly not have a public character.
Furthermore, the Dutch legislator argues that the digital infrastructure is of such importance that attacks on the availability of information systems should almost always be unlawful.

We learned that an effective way of creating the right to digital assemblies can be found in granting the right to manifestation, which is currently only granted by the Dutch constitution\pagenote{\textdutch{Artikel 9 lid 1 van de Grondwet}} and thus not invocable in court\pagenote{\textdutch{Artikel 120 van de Grondwet}}.

Finally, we discussed the requirements that should be met to conform to the definition of a digital assembly -- visibility, expression of opinion, and collectivity --, which resulted in the technical requirements of visibility, one man one vote, and group proportionality.
We also noted requirements for a digital assembly to be permissible, namely a lack of coercion, proportionality, and subsidiarity.
Finally, we found that digital assemblies can be held orderly by implementing revocable anonymity, acknowledging decentralisation, and allowing some means of announcement.

\subsection{Further Research}
Some of the threads started in this paper are of interest for future research.
Digital assemblies also have a social component.
In future research, the social influences of a digital assembly and the power this gives to the participants should be studied.
Related to this, the dynamics of group proportionality, as proposed in Section~\ref{sec:techgroupproportionality}, could be evaluated.

The proposed fundamental rights should be further evaluated.
The intersections between the digital assembly and the Budapest Convention on Cybercrime also require further research.
This could be extended to the related directions and regulations of the European Union.

The issue of jurisdiction will continue to be of high importance, especially when it concerns digital topics.
In this case, it should be explored what happens when the European legislator grants its citizens the digital right to assembly and the government of the USA does not grant this.
This question could also be expanded to collaboration between the European Union and the USA in police activities. 

Some legal requirements are not easily translated to technical requirements, due to their nature.
However, this does leave gaps, as shown in the figures in Appendix~\ref{sec:relations}.
Therefore, we recommend further research on this translation, with a focus on those topics that are not suitable for direct translation to technical requirements.
Additionally, several technical topics are worth further pursuance, such as the implementation of revocable anonymity.

\section*{Acknowledgement}
We are very grateful for the supervision, advise, and knowledge given to us by prof.\ mr.\ dr.\ M.\ Hildebrandt.
For his technical advise, we are very thankful to dr.\ J.H.\ Hoepman.

\printnotes
\printbibliography

\appendix

\section{Relations of the Requirements}
\label{sec:relations}
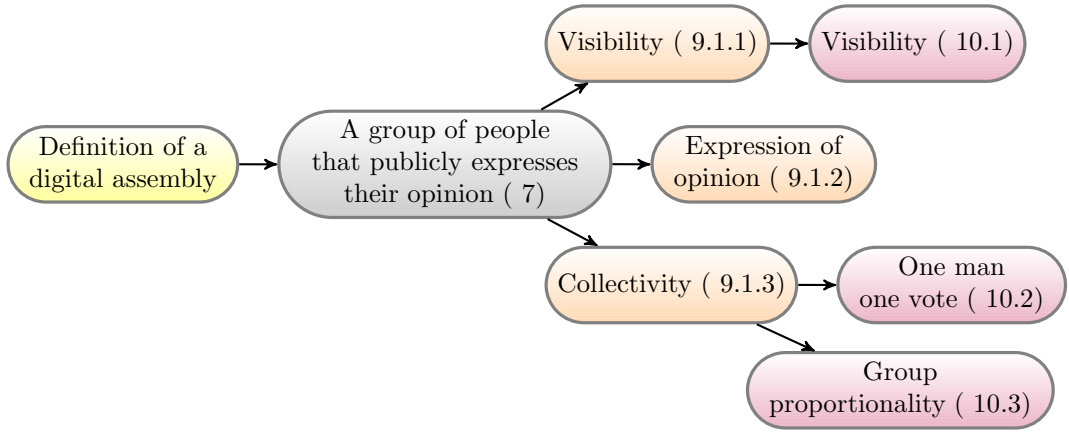
\begin{figure*}
	\centering

	\tikzset{descriptionnode/.style={rounded rectangle,minimum size=10mm,very thick,draw=black!50,top color=white,bottom color=yellow!40}}	
	\tikzset{legalnode/.style={rounded rectangle,minimum size=10mm,very thick,draw=black!50,top color=white,bottom color=black!20}}
	\tikzset{requirenode/.style={rounded rectangle,minimum size=10mm,very thick,draw=black!50,top color=white,bottom color=orange!30}}
	\tikzset{technicalnode/.style={rounded rectangle,minimum size=10mm,very thick,draw=black!50,top color=white,bottom color=purple!30}}
	
	\begin{tikzpicture}[node distance=5mm,thick,>=stealth',align=center,auto]
		\node (description)			[descriptionnode]					{Definition of a\\digital assembly};
		
		\node (definition)			[legalnode,right=of description]			{A group of people\\that publicly expresses\\their opinion (§~\ref{sec:recap})};
		
		\node (visibility)			[requirenode,above right=of definition]		{Visibility (§~\ref{sec:reqvisible})};
		\node (opinion)			[requirenode,right=of definition]			{Expression of\\opinion (§~\ref{sec:reqopinion})};
		\node (collectivity)			[requirenode,below right=of definition]		{Collectivity (§~\ref{sec:reqcollective})};
		
		\node (tvisibility)			[technicalnode,right=of visibility]			{Visibility (§~\ref{sec:techvisible})};
		\node (onemanonevote)		[technicalnode,right=of collectivity]		{One man\\one vote (§~\ref{sec:techonemanonevote})};
		\node (groupproportionality)	[technicalnode,below right=of collectivity]	{Group\\proportionality (§~\ref{sec:techgroupproportionality})};
		
		\draw (description)		edge[->]	(definition);
		
		\draw (definition)		edge[->]	(visibility);
		\draw (definition)		edge[->]	(opinion);
		\draw (definition)		edge[->]	(collectivity);
		
		\draw (visibility)		edge[->]	(tvisibility);
		
		\draw (collectivity)		edge[->]	(onemanonevote);
		\draw (collectivity)		edge[->]	(groupproportionality);
	\end{tikzpicture}
	
	\caption{Relations of the requirements of the proposed definition of a digital assembly.}
	\label{fig:definition}
\end{figure*}

\begin{figure*}
	\centering
	
	\tikzset{descriptionnode/.style={rounded rectangle,minimum size=10mm,very thick,draw=black!50,top color=white,bottom color=yellow!40}}
	\tikzset{legalnode/.style={rounded rectangle,minimum size=10mm,very thick,draw=black!50,top color=white,bottom color=black!20}}
	\tikzset{requirenode/.style={rounded rectangle,minimum size=10mm,very thick,draw=black!50,top color=white,bottom color=orange!30}}
	\tikzset{technicalnode/.style={rounded rectangle,minimum size=10mm,very thick,draw=black!50,top color=white,bottom color=purple!30}}
	
	\begin{tikzpicture}[node distance=5mm,thick,>=stealth',align=center,auto]
		\node (description)			[descriptionnode]						{Permissibility of a\\digital assembly};
		
		\node (defcoercion)			[legalnode,above right=of description]			{The protest is in the\\general interest (§~\ref{sec:recap})};
		\node (defproportionality)		[legalnode,right=of description]				{Possible damages\\are proportional (§~\ref{sec:recap})};
		\node (defsubsidiarity)		[legalnode,below right=of description]			{Alternatives\\have been pursued (§~\ref{sec:recap})};

		\node (opinion)			[requirenode,above right=of defcoercion,dashed]	{Expression of\\opinion (§~\ref{sec:reqopinion})};
		\node (nocoercion)			[requirenode,right=of defcoercion]			{No coercion (§~\ref{sec:reqnocoercion})};		
		\node (proportionality)		[requirenode,right=of defproportionality]		{Proportionality (§~\ref{sec:reqproportionality})};
		\node (subsidiarity)			[requirenode,right=of defsubsidiarity]			{Subsidiarity (§~\ref{sec:reqsubsidiarity})};
		
		\draw (description)		edge[->]	(defcoercion);
		\draw (description)		edge[->]	(defproportionality);
		\draw (description)		edge[->]	(defsubsidiarity);
		
		\draw (defcoercion)		edge[->]	(opinion);
		\draw (defcoercion)		edge[->]	(nocoercion);
		\draw (defproportionality)	edge[->]	(proportionality);
		\draw (defsubsidiarity)	edge[->]	(subsidiarity);
	\end{tikzpicture}
	
	\caption{Relations of the requirements concerning the permissibility of a digital assembly.}
	\label{fig:permissibility}
\end{figure*}
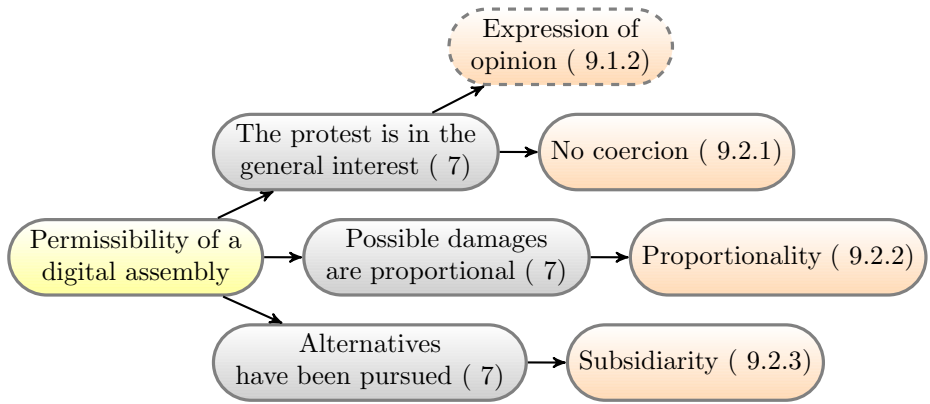

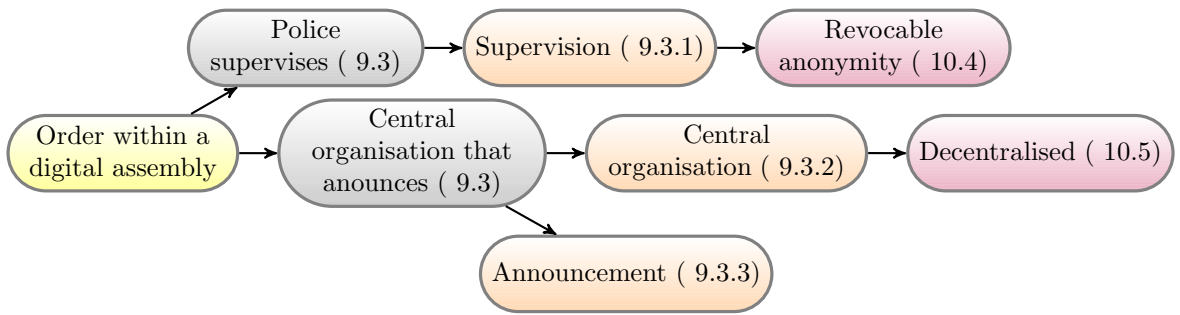
\begin{figure*}
	\centering
	
	\tikzset{descriptionnode/.style={rounded rectangle,minimum size=10mm,very thick,draw=black!50,top color=white,bottom color=yellow!40}}
	\tikzset{legalnode/.style={rounded rectangle,minimum size=10mm,very thick,draw=black!50,top color=white,bottom color=black!20}}
	\tikzset{requirenode/.style={rounded rectangle,minimum size=10mm,very thick,draw=black!50,top color=white,bottom color=orange!30}}
	\tikzset{technicalnode/.style={rounded rectangle,minimum size=10mm,very thick,draw=black!50,top color=white,bottom color=purple!30}}
	
	\begin{tikzpicture}[node distance=5mm,thick,>=stealth',align=center,auto]
		\node (description)			[descriptionnode]						{Order within a\\digital assembly};
		
		\node (police)			[legalnode,above right=of description]			{Police\\supervises (§~\ref{sec:orderly})};
		\node (centralannounce)		[legalnode,right=of description]				{Central\\organisation that\\anounces (§~\ref{sec:orderly})};
		
		\node (supervision)			[requirenode,right=of police]		{Supervision (§~\ref{sec:reqsupervision})};		
		\node (central)			[requirenode,right=of centralannounce]		{Central\\organisation (§~\ref{sec:reqcentral})};
		\node (announcement)		[requirenode,below right=of centralannounce]	{Announcement (§~\ref{sec:reqannouncement})};
		
		\node (anonymity)			[technicalnode,right=of supervision]			{Revocable\\anonymity (§~\ref{sec:techrevocable})};
		\node (decentralised)		[technicalnode,right=of central]				{Decentralised (§~\ref{sec:techdecentral})};
		
		\draw (description)		edge[->]	(police);
		\draw (description)		edge[->]	(centralannounce);
		
		\draw (police)		edge[->]	(supervision);
		\draw (centralannounce)	edge[->]	(central);
		\draw (centralannounce)	edge[->]	(announcement);
		
		\draw (supervision)		edge[->]	(anonymity);
		\draw (central)		edge[->]	(decentralised);
	\end{tikzpicture}
	
	\caption{Relations of the requirements to an orderly a digital assembly.}
	\label{fig:order}
\end{figure*}

In figure~\ref{fig:definition}, the relations of the requirements of the proposed definition of a digital assembly are displayed.
Figure~\ref{fig:permissibility} does the same for the requirements concerning the permissibility of a digital assembly.
Finally, figure~\ref{fig:order} concerns the requirements for orderly digital assemblies.

\end{document}